\documentclass{JCIN18}
\UseRawInputEncoding

\begin{document}

\ArticleType{}
\Year{2021}

\title{A Preamble Based MAC Mechanism in Ad-Hoc Network}

\author[1]{Chenchao Shi}
\author[1]{Biqian Feng}
\author[1]{Yongpeng Wu*}
\author[1]{Wenjun Zhang}

\address[1]{School of Electronic Engineering, Shanghai Jiao Tong University, Shanghai 200240, China}

\abstract{In this paper, we propose a preamble based medium access control (P-MAC) mechanism in Ad-Hoc network. Different from traditional carrier sense multiple access (CSMA) in Ad-Hoc network, P-MAC uses much shorter preamble to establish the network. First, we propose the P-MAC mechanism to shorten the time of establishing the Ad-Hoc network. Based on the P-MAC, we propose a more efficient way to maintain the network. Next, focusing on the power line communication (PLC) network which is a kind of Ad-Hoc network, we propose a frequency division power line communication (FD-PLC) network architecture to obtain the best communication frequency. To obtain the best frequency, i.e., highest SNR, we design the frequency sweeping mechanism which can determine the frequency of uplink and downlink communication before the transmitter and receiver communicate. Due to the large-scale networks in industry, P-MAC can be exploited to speed up the establishment of the Ad-Hoc PLC network. Finally, we compare our mechanism with CSMA. Numerical results indicate that our strategy greatly shortens the time of establishing the Ad-Hoc network.}


\maketitle


\footnote{C. Shi, B. Feng, Y. Wu, W. Zhang are with the Department of Electronic Engineering, Shanghai Jiao Tong University, Minhang 200240, China (e-mail: 624006248@sjtu.edu.cn; fengbiqian@sjtu.edu.cn;
yongpeng.wu@sjtu.edu.cn; zhangwenjun@sjtu.edu.cn) (Corresponding author: Yongpeng Wu).}

\section{Introduction}
\subsection{Background}
A wireless Ad-Hoc network consists of a collection of wireless nodes that are connected without any fixed infrastructure or centralized administration. Each network node can serve as a router that forwards packets for others. A flow from the source node to the destination node may take multiple hops to traverse over multiple wireless links. Although IEEE 802.11 protocol is widely used in the wireless Ad-Hoc networks, most studies focus on end-to-end throughput optimization \cite{end-to-end throughput}, energy-efficient protocol \cite{energy-efficient protocol} but ignore the work related to reducing time complexity of establishing network.

The traditional centralized multi-hop Ad-Hoc network medium access control (MAC) protocol is based on the carrier sense multiple access with collision avoid (CSMA/CA) competition mechanism \cite{CSMA}, forming a tree network topology and allowing backup routes to improve the stability of network transmission. However, this type of access method uses data transmission for competitive access \cite{IEEE MAC}. Therefore, the time complexity of establishing and maintaining network is high when the number of networks is large.

Ad-Hoc networks are now widely used in power line communication (PLC) \cite{PLC}. Focusing on the traditional PLC network, there are three critical channel  parameters, namely, noise, impedance and attenuation, which are found to be highly unpredictable and fluctuant over time, frequency and location. In order to overcome these difficulties, a lot of efforts have been undertaken to characterize a power line channel model [7]-[9]. Due to the complexity and diversity of the channel, it is difficult for the traditional PLC to guarantee the validity of the traffic on a fixed frequency.

\subsection{Our work}

The intention of our work is to improve the efficiency of CSMA networking and the reliability of PLC transmission. Therefore, we first propose a preamble based medium access control mechanism (P-MAC) in Ad-Hoc network. Different from traditional CSMA in Ad-Hoc network,  P-MAC uses much shorter preamble to access and establish the network. A preamble time exchange process is required to record the time difference between receiving and transmitting the preamble. The P-MAC reduces the time complexity of establishing network by approximately 60 times compared with traditional CSMA access as mentioned in Section 4. Secondly, since the optimal communication frequency of the PLC channel at various locations is not fixed \cite{PLC Channel Characteritics}, we propose a frequency division power line communication (FD-PLC) network architecture to obtain the best communication frequency. We design the frequency sweeping mechanism which can determine the frequency of uplink and downlink communication before the transmitter and receiver communicate. And the frequency sweeping mechanism only requires $(2N+1)$ communications, where $N$ represents the number of the up-link and downlink frequency points of the PLC. Next, we attempt to incorporate P-MAC mechanism into FD-PLC to speed up the establishment of the Ad-Hoc PLC network.

In general, the contributions of our work are summarized as follows:

\begin{itemize}
\item P-MAC is proposed to shorten the time of establishing the Ad-Hoc network. P-MAC utilizes the preamble time exchange (PTE) process to record the time difference between receiving the preamble and transmitting the preamble. This time difference allows the transmitter and receiver to communicate with each other to confirm their identity.We design different preambles in order to distinguish the PTE mode and the data mode. The comprehensive agreement mechanism consists of PTE process, T-Query handshake, and network configuration (Net-Config) handshake. P-MAC replaces data transmission with preamble frames for competitive access so that it greatly reduces the time of establishing the Ad-Hoc network.

\item Based on the P-MAC, we propose how to maintain the network. The node notifies the gateway through the method of reporting layer by layer to extend the survival period of the node.

\item Unlike wireless and proprietary communication line technologies, PLC channels vary greatly with the power line topology, access to electrical equipment and communication locations. It is difficult to select an effective communication frequency band, and the optimal communication frequency band is often time-varying. The existing PLC standards and technologies are mostly based on fixed communication frequency bands. Therefore, we propose a FD-PLC architecture which enables PLC to communicate with multiple frequencies. 

\item Based on thr FD-PLC architecture, we desigin a frequency division single input single output PLC (SISO-PLC) system, and configure the digital front-end on the transmitter and receiver as multiple channels in order to formulate a frequency sweeping mechanism suitable for PLC channels with a shorter frequency sweeping cycle. Compared with the traditional PLC system, the frequency division system can not only realize the configurable frequency point and bandwidth, but also realize the transmission of multiple frequency domain incoherent carrier signals and the reception of any frequency point carrier signal. Then the frequency sweeping mechanism in the frequency division system is designed. The frequency sweeping mechanism only requires $2N+1$ times of communication so that the frequency sweeping efficiency is greatly improved.

\item The combination of P-MAC and FD-PLC makes the PLC system capable of efficient multi-frequency establishing network.
\end{itemize}

The P-MAC can be used in other communication networks which use CSMA to establish the network. It not only focused on power line communication networks and other ad hoc network such as chain network, star network and wheel network.

The rest of this paper is organized as follows. In Section \uppercase\expandafter{\romannumeral2}, we propose our P-MAC access mechanism. In Section \uppercase\expandafter{\romannumeral3}, we propose the FD-PLC architecture. Numerical results are provided in Section \uppercase\expandafter{\romannumeral4}. Finally, conclusions are drawn in Section \uppercase\expandafter{\romannumeral5}.

\section{Preamble Based MAC Mechanism}
The traditional centralized multi-hop Ad-Hoc network MAC protocol \cite{AdHoc} is based on the CSMA/CA competition mechanism, forming a tree network topology and allowing backup routes to improve the stability of network transmission. However, this type of access method uses data transmission for competitive access so that the time complexity of establishing network is too high. To handle the difficulty, a P-MAC competitive access mechanism is proposed to replace the data transmission with preamble frames for competitive access, which greatly reduces network establishment time.

\subsection{Comprehensive Agreement Mechanism}
The P-MAC mechanism utilizes the preamble time exchange (PTE) process to record the time difference between receiving and transmitting the preamble. In the PTE process, the gateway node sends the preamble in the form of broadcast. The nodes that want to access the network reply the preamble by competitive access and record the time difference between receiving and transmitting the preamble. In the case of effective preamble transmission, this time difference information can be known at both the transmitting and the receiving node. The corresponding time consumption is related to the length of the preamble frame and the competitive access progress.

\begin{figure}[!t]
	\centering
	\includegraphics[width=0.5\textwidth]{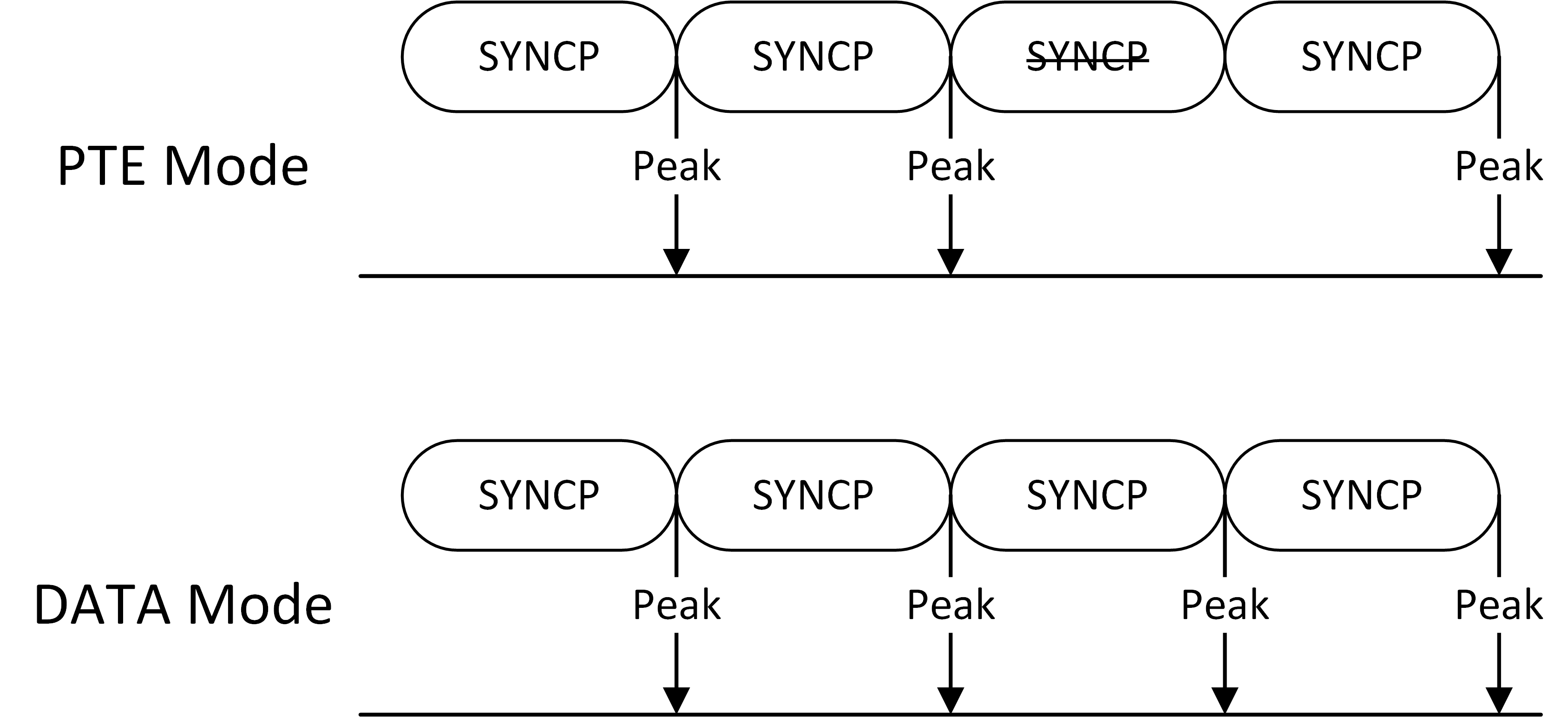}\\
	\caption{Comparison of PTE mode based on missing SYNCP and DATA mode}
	\label{Figure_1}
\end{figure}

The preamble design of PTE uses different combinations of synchronous cyclic prefix (SYNCP) \cite{SYNCP} to determine whether the frame is in PTE mode or DATA mode. As shown in Fig. \ref{Figure_1}, a feasible design to distinguish two modes is that the PTE preamble uses three SYNCPs separated by a SYNCP empty position while the data transmission preamble uses four consecutive SYNCPs. Through the cross-correlation calculation in the frame detection, the SYNCP generates a peak. The absence of SYNCP detected indicates the PTE mode. Therefore, the mode is determined by judging the change position. In dealing with different scenarios, different changes can be made to the frame structure to satisfy different needs.

After PTE process, the T-Query handshake is used to determine the node information of exchange time. The node that originally sent the preamble sends out a data packet containing time difference information. The node which received this data packet compares with the locally stored time difference information. If the time difference is the same as the locally stored time difference, the node's MAC address or short identity (SID) address is replied to the transmitting node. 

\begin{figure*}[!t]
	\centering
	\includegraphics[width=1.0\textwidth]{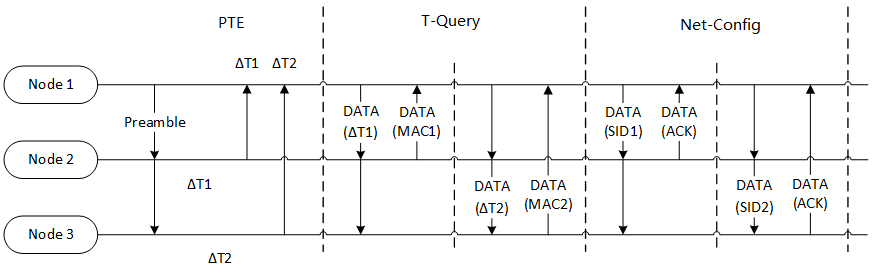}\\
	\caption{Basic structure of the establishment process}
	\label{Figure_2}
\end{figure*}

In addition, the Net-Config handshake is designed to allocate the SID address according to the MAC address, and confirm whether the node is connected to the network. If it has been connected, the node does not respond in the PTE process and replies the preamble. In the T-Query handshake process, it means that the node has not been assigned an SID address if the node replies with a MAC address. In the Net-Config handshake process, the transmitting node must send the corresponding SID address, and the receiving node will bind the SID address after receiving it. Next is the explanation of the basic structure of the establishment process.

\textbf{PTE:} As shown in Fig. 2, node 1 is the network gateway and both node 2 and 3 want to access the network. According to the access procedure above, node 1 sends the preamble. Nodes 2 and 3 record the time difference between receiving and transmitting the preamble. The time difference recorded by node 2 denotes  $\Delta$T1 and that of node 3 denotes $\Delta$T2. After node 1 receives the reply preamble, it also records the time difference between transmitting and receiving the preamble including $\Delta$T1 and  $\Delta$T2. 

\textbf{T-Query handshake:} Node 1 sends a data packet with time difference information  $\Delta$T1. After nodes 2 and 3 receive this time difference, they compare with the locally stored time difference information. Node 2 sends back the data packet which contains the MAC address of node 2 (MAC1). Node 1 continues to send data packets with time difference information  $\Delta$T2. After nodes 2 and 3 receive this time difference, they compare with the locally stored time difference information. Node 3 sends back the data packet which contains the MAC address of node 3 (MAC2).

\textbf{Net-Config handshake:} According to the received MAC address, node 1 queries the SID and MAC binding table, and sends a data packet to node 2, including MAC1 and SID1. After receiving it, node 2 sets the local SID and responds with confirmation. According to the received MAC address, node 1 queries the SID and MAC binding table, and sends a data packet to node 3, including MAC2 and SID2. After receiving it, node 3 sets the local SID and responds with confirmation.

\subsection{Implement of P-MAC Mechanism}

For multi-level networks, the network gateway performs the P beacon to reach the target node through the relay node. The PTE process is performed by the target node. At the end of this process, the target node returns the time difference information to the gateway through the relay node. The gateway then performs a T-Query handshake based on the time difference information. At this time, there are two ways to access. i) If fair access is required,  all target nodes wait to complete the PTE and T-Query handshake. Then the network starts the Net-Config handshake for all target nodes, and at the same time assigns SID addresses, clarifying which target nodes are willing to connect to the network. ii) If fast access is required, the Net-Config handshake process can be directly performed after the PTE and the T-Query handshake process which assigns the SID address and clarifies whether to access or not.

Based on this design, the P-MAC access process can be clarified. The gateway completes the access of the first-level node based on the PTE process, T-Query process, and Net-Config process. After that, the gateway completes the access of the second-level node according to the first-level node network through P beacon, relay, PTE, T-Query handshake, and Net-Config handshake, and builds the second-level node network according. The construction of three-level and multi-level networks is the same as the second-level which is also completed by the gateway according to the upper-level node network through P beacon, relay, PTE, T-Query handshake, and Net-Config handshake.

\begin{figure*}[!t]
	\centering
	\includegraphics[width=1.0\textwidth]{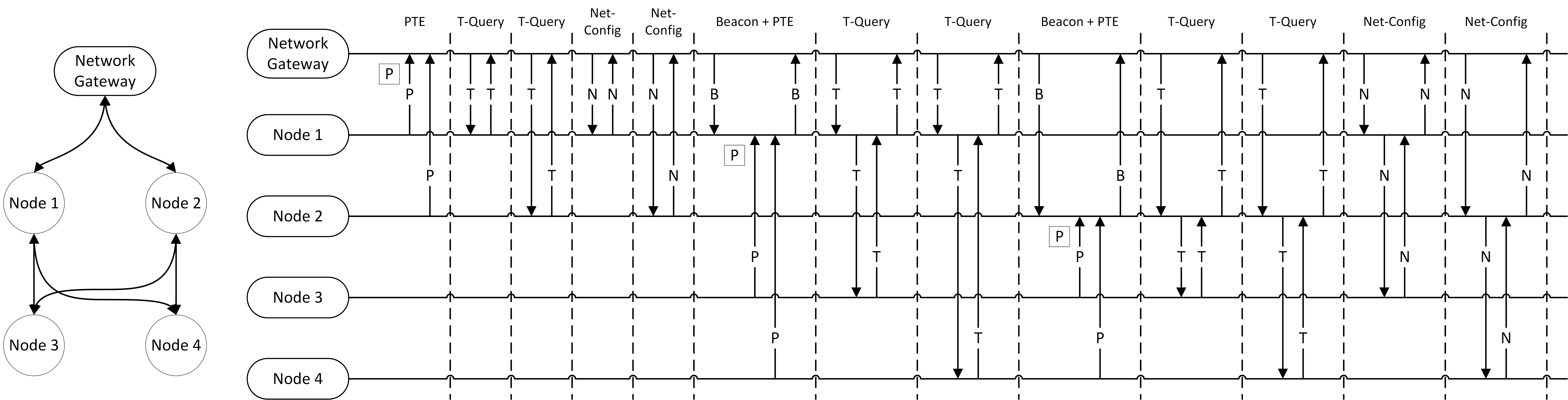}\\
	\caption{Establishment of a two-level network}
	\label{Figure_3}
\end{figure*}

The following analyzes an example of a two-level network composed of 4 nodes. A topological graph is leveraged to describe the network link status in Fig. 3. The gateway can communicate with nodes 1 and 2. Both node 1 and node 2 can communicate with nodes 3 and 4. The network establishment process is performed by the gateway. The gateway performs PTE, and nodes 1 and 2 reply the preamble. The gateway performs a T-Query handshake process and sends the time difference information. And then the gateway matches with node 1 and obtains the MAC address. The gateway continues to perform a T-Query handshake transmitting the time difference information, matching with node 2 and obtaining the MAC address. The gateway performs the Net-Config handshake according to the channel quality in the T-Query handshake process, sends the bound SID address and network ID (NWKID) to confirm the network access through the MAC address of node 1. Node 1 replies confirmation. The gateway continues to perform the Net-Config handshake, and sends the bound SID address and NWKID through the MAC address of node 2 to confirm access to the network. Node 2 replies confirmation. Up to this point, the construction of the first level network is completed.

The gateway performs a P beacon to node 1. Node 1 performs a PTE after receiving it. Nodes 3 and 4 reply the preamble, and node 1 returns the time difference information to the gateway. The gateway performs a T-Query handshake and node 1 performs a T-Query handshake after receiving it, transmitting the time difference information, matching with node 3 and obtaining the MAC address which is replied to the gateway. The gateway continues to perform a T-Query handshake. After receiving it, node 1 performs a T-Query handshake and sends the time difference information to match with node 4 and to obtain the MAC address which is  replied to the gateway.

The gateway performs a P beacon to node 2. Node 2 performs a PTE after receiving it. Nodes 3 and 4 reply the preamble, and node 2 returns the time difference information to the gateway. The gateway performs a T-Query handshake and node 2 performs a T-Query handshake after receiving it, transmitting the time difference information, matching with node 3 and obtaining the MAC address which is replied to the gateway. The gateway continues to perform a T-Query handshake. After receiving it, node 2 performs a T-Query handshake, sends the time difference information to match with node 4 and to obtain the MAC address which is  replied to the gateway.

The gateway selects route 1 as the communication route based on the two communication channel quality with node 3, i.e., route 1 : gateway $\Rightarrow$ node 1 $\Rightarrow$ node 3; route 2 : gateway $\Rightarrow$ node 2 $\Rightarrow$ node 3. The gateway performs a Net-Config handshake to node 3 through node 1, and sends the bound SID address and NWKID to confirm network access through the MAC address of node 3. Node 3 replies confirmation. The gateway selects route 2 as the communication route according to the quality of the communication channel with node 4 twice, i.e., route 1: gateway $\Rightarrow$ node 1 $\Rightarrow$ node 4; route 2: gateway $\Rightarrow$ node 2 $\Rightarrow$ node 4. The gateway performs a Net-Config handshake to node 4 through node 2 and sends the bound SID address and NWKID to confirm network access through the MAC address of node 4. Node 4 replies confirmation. Up to this point, the construction of the second-level network is completed and the two-level network is established. More levels of the network can also be expanded in the same way. 

\subsection{Network maintenance process}
The following explains the maintenance process of the network based on the P-MAC protocol according to the three-level network. For the first-level network, the gateway initiates a PTE. After the node receives the preamble, the preamble competition response is not performed if it is in the network-connected state. Gateway initiates a Net-Config handshake to node 1 according to the previous network access list, confirms the target node through SID, configures the heartbeat period and NWKID, node 1 responds to confirmation after receiving it, and updates the survival period, i.e., set to the heartbeat period. The maintenance process of node 2 is the same as that of node 1. If node 2 exceeds the heartbeat period and does not receive the T-Query handshake, it will automatically exit the network but retain the SID binding. At this time, gateway initiates PTE, and node 2 reply to the preamble and save the time difference information. The gateway then initiates a T-Query handshake based on this time difference information to obtain the information of node 2. The gateway then connects node 2 to the network through the Net-Config handshake after traversing all secondary nodes.

For the second-level network, gateway initiates a P beacon to node1. Node 1 initiates PTE after receiving the P beacon, node 3 and node 4 receive the preamble, and if they are still in the network connection state, they do not reply to the preamble. Gateway continues to initiate P beacon to node 2, and node 2 initiates PTE after receiving the P beacon. Similarly, node 3 and node 4 receive the preamble, and if they are still in the network connection state, they do not reply to the preamble. Considering that ndoe 3 has left the network, it replies to the preamble. Two PTEs generate two T-Query to obtain node 3 information. Since the SID has been allocated, node 3 returns the SID address information. After that, the gateway accesses it through the Net-Config handshake according to the returned channel quality. If there is no leading reply in the process, the gateway will perform the T-Query handshake one by one according to the previous network access list and the feasible route to extend the node's survival period.

For the third-level network, the maintenance process is consistent with the second-level network, except that the gateway initiates the P beacon, PTE, T-Query handshake, and Net-Config handshake all through the relay route.

\begin{figure*}[!t]
	\centering
	\includegraphics[width=1.0\textwidth]{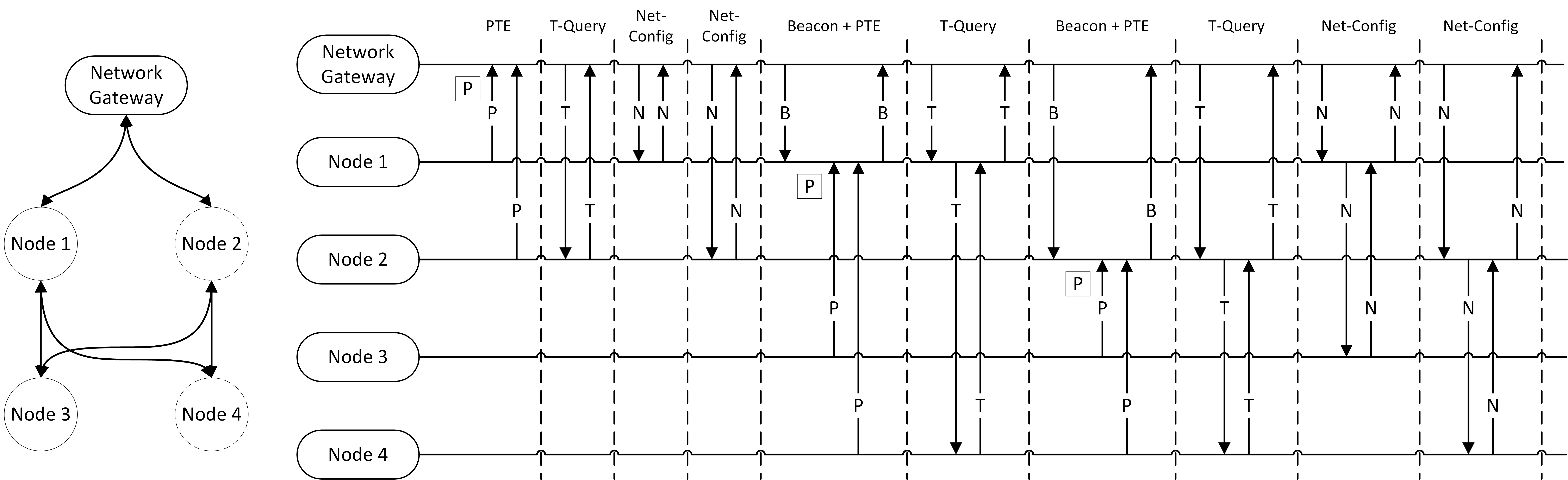}\\
	\caption{Maintenance process of a two-level network}
	\label{Figure_3}
\end{figure*}

As shown in the Fig. 4, according to the above network maintenance process, the following analyze how to maintain the network after losing nodes 2 and 4 in the above-mentioned network, while reconnecting to the lost nodes. The network maintenance process is initiated by the gateway. The gateway initiates PTE and node 2 replies to the preamble. Node 1 does not participate in the PTE because it has already connected to the network. The gateway initiates a T-Query handshake and sends time difference information to matche with node 2 and to obtain the MAC address. The gateway initiates an Net-Config handshake to node 1 according to the list of connected nodes to extend the survival period of node 1. Node 1 replies with confirmation. The gateway initiates the Net-Config handshake according to the channel quality in the T-Query handshake phase and sends the bound SID address and NWKID to confirm the network access through the MAC address of node 2. Node 2 replies to confirm. Hereby, the maintenance of the first level network is completed.

The gateway initiates a P beacon. Node 1 initiates a PTE after receiving it, node 4 replies to the preamble, and node 1 replies the time difference information to the gateway. Node 3 does not participate in PTE because it has already connected to the network. The gateway initiates a T-Query handshake. Node 1 initiates a T-Query handshake after receiving it and sends the time difference information to match with node 4. Then the gateway obtains the node 4's MAC address. After that, node 1 replies to the gateway.

The gateway initiates a P beacon. Node 2 initiates a PTE after receiving it, node 4 replies to the preamble, and node 2 replies the time difference information to the gateway. Node 3 does not participate in PTE because it has already connected to the network. The gateway initiates a T-Query handshake. Node 2 initiates a T-Query handshake after receiving it and sends the time difference information to match with node 4. Then the gateway obtains the node 4's MAC address. After that, node 2 replies to the gateway.

The gateway initiates a Net-Config handshake to node 3 through node 1 according to the list of nodes that have already connected to the network to extend the survival period of node 3. After that, node 3 replies with confirmation. The gateway selects route 2 as the communication route according to the quality of the two communication channel with node 4, i.e., route 1: gateway $\Rightarrow$ node 1 $\Rightarrow$ node 4; route 2: gateway $\Rightarrow$ node 2 $\Rightarrow$ node 4. The gateway initiates a Net-Config handshake to node 4 through node 2 and sends out the bound SID address and NWKID to confirm network access through the MAC address of node 4. After that, node 4 replies with confirmation.

Up to this point, the maintenance of the second-level network is completed.

\section{Frequency Division Power Line Communication Network}
Unlike wireless and proprietary communication line technologies, PLC channels vary greatly with the power line topology, access to electrical equipment and communication locations. In general, the PLC channel has properties of frequency selectivity, time-varying, colored background noise, narrow-band interference, and various impulse noises. The environment noise of the low frequency band is much higher than that of the high frequency band. In the low frequency band and the high frequency band, the signal attenuation exhibits great frequency selectivity, which is not highly correlated with the communication distance. Observing the PLC channel from the perspective of low frequency band and high frequency band, it is difficult to select an effective communication frequency band, and the optimal communication frequency band is often time-varying.

The existing PLC standards and technologies are mostly based on fixed communication frequency bands, narrowband (90 kHz - 500 kHz) and broadband (2 MHz - 30 MHz), and use repetitive coding to counter the frequency selectivity and color of PLC channels. Background noise interference causes communication rate problems. Other PLC technologies adapt to the PLC channel by configurable center frequency and bandwidth. The transmitter and receiver need to implement a complex frequency sweeping mechanism to negotiate the center frequency and bandwidth of the uplink and downlink. In the case of constant bandwidth, a typical sweep time is $N^{2}+N$ communications where $N$ represents the number of center frequency configurations. In addition, since the receiver can monitor only one type of center frequency and bandwidth configuration, the active event reporting of the lower computer cannot be realized, and it can only be polled by the upper computer.

\subsection{Frequency Division Power Line Communication}
Most of the existing wireless communication technologies support frequency division multiplexing technology to avoid edge cross-talk between cells. Since the channel characteristics of the adjacent frequencies of the wireless communication channel do not change much, the wireless communication technology is negotiated based on a fixed frequency and then the entire cell is migrated to the new frequency to achieve frequency division multiplexing. However, a fixed frequency band cannot be obtained in the PLC channel for negotiation, so it is necessary to consider a frequency division multiplexing mechanism that is suitable for the PLC channel.

The digital front-end in the form of equivalent complex baseband is widely used in the field of wireless communication. The bandwidth of the digital signal can be changed by configuring the number of interpolation and decimation filter stages. The center frequency of the digital signal can be changed by configuring the frequency of the mixer. The application of the equivalent complex baseband digital front-end in PLC is not implemented in multiple channels on the receiver. Only one channel is implemented in the transmitter and the receiver respectively.

The structure and the principle of the digital front-end motivate us to realize the adjustable frequency and bandwidth of the transmitted signal, which is of great significance to cope with the PLC frequency selectivity. After realizing that the central frequency of the transmitted signal can be adjusted, a new problem is that the transmitter and the receiver should communicate specifically at which frequency in the frequency band of PLC. Next, how to determine the best frequency point for uplink and downlink communication will be introduced.

A frequency sweeping mechanism is required to determine the frequency of uplink and downlink communication before the transmitter and receiver communicate. In a typical PLC system, the upper computer needs to communicate $N$ times in each round, and then receives the signal sent by the lower computer once. At the same time, the lower computer also needs to switch at $N$ frequency points, so the typical sweep time is $N^{2}+N$ times of communication. It can be seen that a typical frequency sweeping time requires $N^{2}+N$ times of communication under the condition of the same bandwidth in order to determine the best uplink and downlink communication frequency point. For the incomplete frequency sweeping mechanism which only requires finding the frequency points where the uplink and the downlink can communicate, the typical frequency sweeping time requires $N/2\cdot(N+1)+2$ times of communications where $N$ represents the number of the uplink and downlink frequency points of the PLC. $N$ is also the number of digital front-end center frequency point configurations. Such an excessive frequency sweeping overhead will seriously affect the communication efficiency.

There are two main reasons for the long sweep period of the traditional PLC system. One is that its receiver can only monitor a central frequency and bandwidth configuration, and cannot realize active event reporting from the lower computer. The events can only be polled by the upper computer. On the other hand, unlike wireless communication technology based on a fixed frequency for negotiation, a fixed frequency band cannot be defined for negotiation in the PLC channel.

In this case, this section designs the frequency division SISO-PLC system, and configures the digital front-end on the transmitter and receiver as multiple channels in order to formulate a frequency sweeping mechanism suitable for PLC channels with a shorter frequency sweeping cycle. Compared with the traditional PLC system, the frequency division system can not only realize the configurable frequency point and bandwidth, but also realize the transmission of multiple frequency domain incoherent carrier signals and the reception of any frequency point carrier signal. Then the frequency sweeping mechanism under the frequency division system is designed. The frequency sweeping mechanism only requires ($2N+1$) times of communication so that the frequency sweeping efficiency is greatly improved.

\subsection{Frequency Division SISO-PLC System}
\begin{figure*}[!t]
	\centering
	\includegraphics[width=1.0\textwidth]{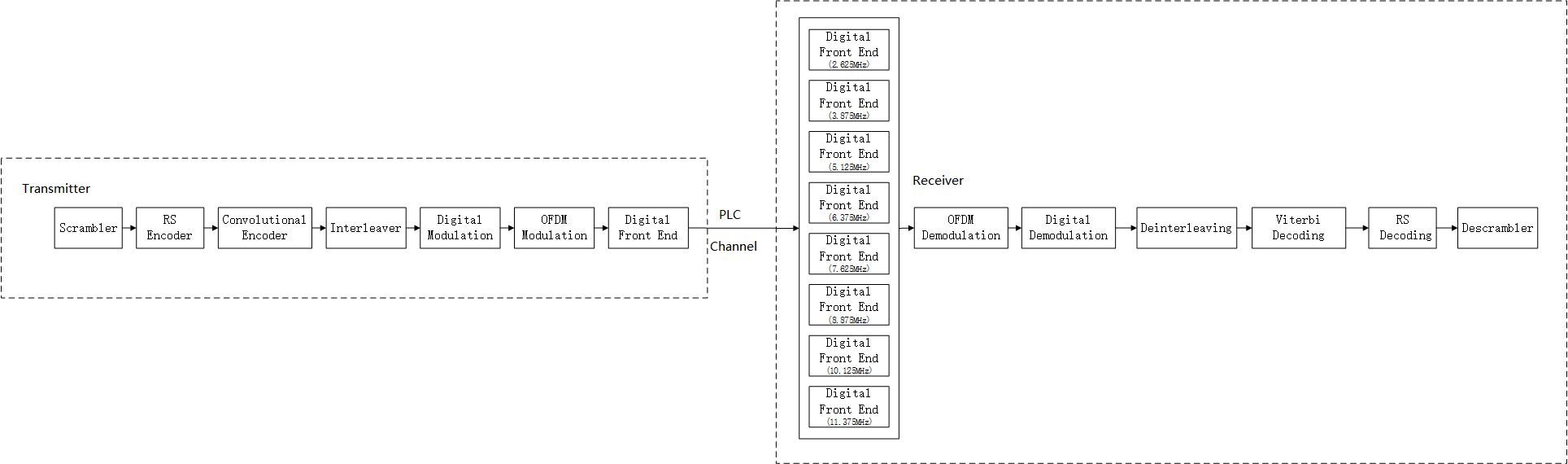}\\
	\caption{System block diagram of frequency division based PLC system.}
	\label{Figure_4}
\end{figure*}

Fig. 5 shows a block diagram of a frequency division SISO-PLC system. The characteristic of this frequency division PLC system is that the digital front-end parts of the transmitter and receiver are configured in a multi-channel form. Among them, the digital front-ends of each channel are configured with different central frequency points which can realize the multi-frequency point transmission of the transmitter and the reception of any frequency point of the receiver. And the center frequency of the digital front-end in the transmitter and the receiver are the same. It ensures that the carrier signal can be correctly received and demodulated.

The data transmitting and receiving process in the Fig. 5 can be simply described as follows: On the transmitter side, the bit stream transmitted by the data link layer passes through the scrambler, RS encoder, convolutional encoder, interleaver, digital modulation and OFDM modulation. After modulation, it is sent to the digital front-end to adjust the center frequency and bandwidth. The bit stream is converted from a digital baseband signal to a digital bandpass signal, and then sent to the power line for transmission. Before data communication, the frequency sweeping operation of the PLC channel can be performed. Through the frequency sweeping mechanism designed in the next section, the uplink and downlink communication frequency with the best signal-to-noise ratio can be selected among the $N$ frequency points that can be communicated. During data communication, the digital front-end is used to adjust the carrier signal to the best frequency point for transmission, so as to effectively avoid the frequency selective interference of the PLC channel or the noise interference on the spectrum.

On the receiving side, the multi-channel digital front-end can perform parallel detection of multiple frequency domain incoherent carrier signals with the same bandwidth and different center frequencies so as to achieve blind detection of carrier signals at any frequency of the transmitter. The multi-channel digital front-end configuration parameters in the receiver correspond to the multi-channel digital front-end parameters in the transmitter one-to-one, so it can ensure that the transmitted carrier signal is correctly received and demodulated. The multi-stage decimation filter inside the digital front-end performs down-sampling processing on the carrier signal, and then converts the digital bandpass signal into a digital baseband signal through a mixer. And then the digital front-end performs OFDM demodulation, digital demodulation, de-interleaving, Viterbi decoding, RS decoding, descrambling and other operations in sequence to demodulate the original bit stream data.

\subsection{Frequency Sweeping Mechanism Under Frequency Division System}

For the frequency division SISO-PLC system, the center frequency is configurable and the bandwidth is configurable so the transmitter can send signals at any frequency and the receiver can receive signals at any frequency. A sweeping mechanism is shown in Fig. 6.
\begin{figure}[!t]
	\centering
	\includegraphics[width=0.5\textwidth]{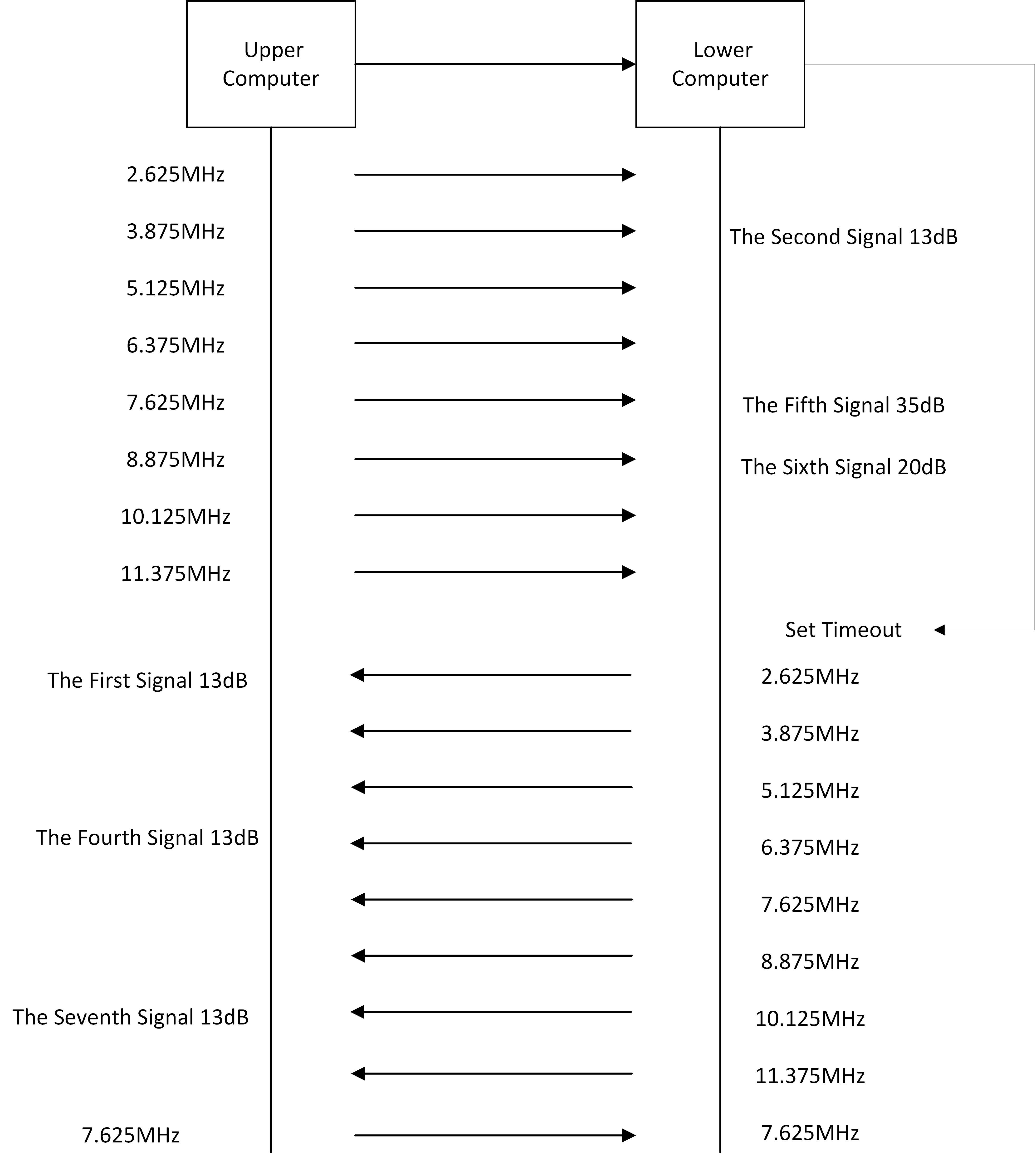}\\
	\caption{Sweep mechanism of frequency division based PLC system}
	\label{Figure_5}
\end{figure}

In Fig. 6, the upper and lower computer represent the sending and receiving end, respectively. When sweeping the frequency to determine the best uplink and downlink frequency, the transmitter and receiver are integrated in the communication equipment. Therefore, when the communication equipment receives a carrier signal with a specific center frequency and bandwidth, it can also send another configuration at the same time which can speed up the frequency sweeping.

The frequency sweeping process in Fig. 6 can be described as follows:

1) The upper computer sends signals successively at $N$ frequency points where $N$ = 8 is selected and the frequency band is 2-12MHz;

2) After the receiver receives the signal, it determines the frequency of the signal and records the SNR of the downlink frequency band this time and then transmits a signal back at this frequency;

3) After the transmitter receives the signal, it records the SNR of this uplink frequency band;

4) After the upper computer and the lower computer have separately recorded the SNR of all $N$ frequency points, the best frequency points for uplink and downlink are determined by comparison;

Due to the characteristics of the frequency division system, it is only necessary for the transmitting end to send carrier signals $N$ times in sequence and the receiving end to send carrier signals $N$ times so that the best uplink and downlink communication frequency can be obtained. Therefore, the entire frequency sweeping time is (2N+1) communications.

\begin{table*}[!t]
	\belowrulesep=0pt 
	\aboverulesep=0pt 
	\renewcommand{\arraystretch}{1.4} 
	\doublerulesep 2.2pt
	\caption{\label{Table_1} Communication time under different frequency sweeping mechanisms.}
	\footnotesize
	\begin{tabularx}{\textwidth}{@{\hspace*{3pt}}@{\extracolsep{\fill}}cXX@{\hspace*{3pt}}}
		\toprule
Pattern & \hspace{4cm}Number of Communications \ \\\hline
Traditional PLC System N×N Frequency Sweeping Mechanism & \hspace{5cm}$N^{2}+N$
\\
Traditional PLC System Incomplete Frequency Sweeping Mechanism    & \hspace{4.5cm}$N/2\cdot(N+1)+2$
\\
Frequency Division PLC System Frequency Sweeping Mechanism
& \hspace{5cm}$2N+1$
\\
\bottomrule
	\end{tabularx}
\end{table*}

The difference from the frequency sweeping mechanism supported by the traditional PLC system is that the frequency division PLC system takes less time to complete the frequency sweeping to determine the best communication frequency point. Under the traditional PLC system, the typical time required to complete a frequency sweeping is ($N^{2}+N$) communications, and the typical time to complete an incomplete frequency sweeping mechanism is ($N/2\cdot(N+1)+2$) communications. Since only one digital front-end is configured in the traditional PLC system, when the next communication frequency point is sent for testing, the internal parameters of the digital front-end need to be re-adjusted, which will inevitably bring time requirements beyond the number of communications required by the frequency sweeping mechanism. For frequency division PLC systems, the receiver is equipped with multiple digital front-ends so that it have the ability to blindly detect carrier signals at any frequency point. It does not need to wait for the polling of the host computer and it can feed back to the host in time after receiving the signal. In short, the structural design of the frequency division system brings great convenience to the system frequency sweeping mechanism.

\subsection{Frequency Division Power Line Network System}

The existing centralized networks are all based on the single-frequency mode. Their communication signals are all on a fixed frequency band. Centralized networks based on the multi-frequency mode all use one of the frequency bands as the main communication mode, and the other frequency bands are used for resource allocation. The centralized network in single frequency mode only needs the beacon mechanism and CSMA time slot to realize the dynamic access of nodes. The mechanism is shown in the Fig. 7 below.

\begin{figure*}[!t]
	\centering
	\includegraphics[width=1.0\textwidth]{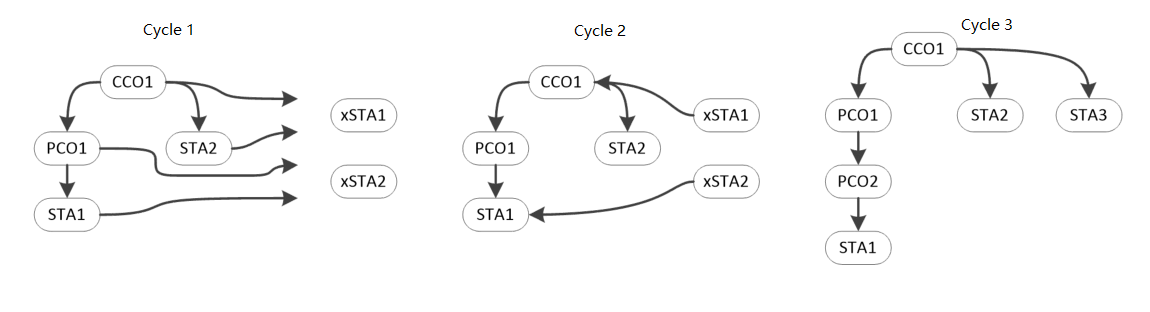}\\
	\caption{Centralized network networking in single frequency mode}
	\label{Figure_6}
\end{figure*}

The nodes Central Control Operator (CCO), Proxy Coordination Operator (PCO), and Station (STA) that have joined the network send beacon frames to the outside in the beacon slot. The node Un-Associated Station (xSTA) that has not joined the network receives the beacon frame in the beacon slot and judges the best access node according to the relevant algorithm. In the CSMA time slot, xSTA1 chooses to access CCO1, and xSTA2 chooses to access STA1. After transmitting the association request and getting the association request reply, STA1 is upgraded to PCO2. xSTA2 is upgraded to STA1 and xSTA2 is upgraded to STA3.

During the access process, the association request establishs the required routing table for each node. After completing the networking, the CCO can arrange TDMA time slots to communicate with the PCO or STA as needed, or receive active reports from the PCO or STA. 

In order to solve the problem of multi-frequency access and multi-frequency two-way communication between nodes, the following networking mechanism is proposed.

As shown in Fig. 8, when the network is started, the Management Node (MN) sends the beacon frame in a broadcast mode in the first cycle and obtains the xLeaf Node (xLN) of the beacon frame to make a registration and association request. After being approved by the MN, the MN will assign a short address to xLN and promote its role to Leaf Node (LN). In the second cycle, MN and LN send the beacon frame in broadcast mode, obtain the xLN of the beacon frame, and select MN or a suitable LN to access and send registration and association requests. If it is to access the MN, after being approved, the MN will assign a short address to xLN and promote its role to LN. If it is to access LN, LN will make a request to MN. After obtaining MN's approval, MN assigns a short address to xLN. The role of LN is now promoted to Relay Node (RN), and the role of xLN is promoted to LN. In the third cycle, the MN, RN, and LN send the beacon frame in a broadcast mode, and the xLN that obtains the beacon frame selects the MN or a suitable RN and LN for access, and sends registration and association requests. If it is to access the MN, after being approved, the MN will assign a short address to xLN and promote its role to LN. If it is to access LN, after approval, LN will make a request to MN. After being approved by the MN, the MN assigns a short address to xLN. The role of LN is now promoted to RN, and the role of xLN is promoted to LN. If it is to access the RN, the RN will make a request to the MN after being approved. After obtaining MN approval, MN assigns a short address to xLN. The role of xLN is promoted to LN. The following cycles all operate in accordance with the mechanism of the third cycle. Data communication in the network is divided into two parts: uplink and downlink and data can be exchanged among MN, RN, and LN.

Since large number of nodes exist in the industrial PLC network, the FD-PLC network and P-MAC can be combined to achieve the efficient multi-frequency networking. P-MAC replaces data transmission with preamble frames for competitive access so that it greatly reduces the time of establishing the PLC network. The combination of FD-PLC and P-MAC can improve the traditional power line networking time by two orders of magnitude, and it can be networked at multiple frequencies. The specific effect will be verified by experimental simulation in the next section.

\begin{figure*}[!t]
	\centering
	\includegraphics[width=1.0\textwidth]{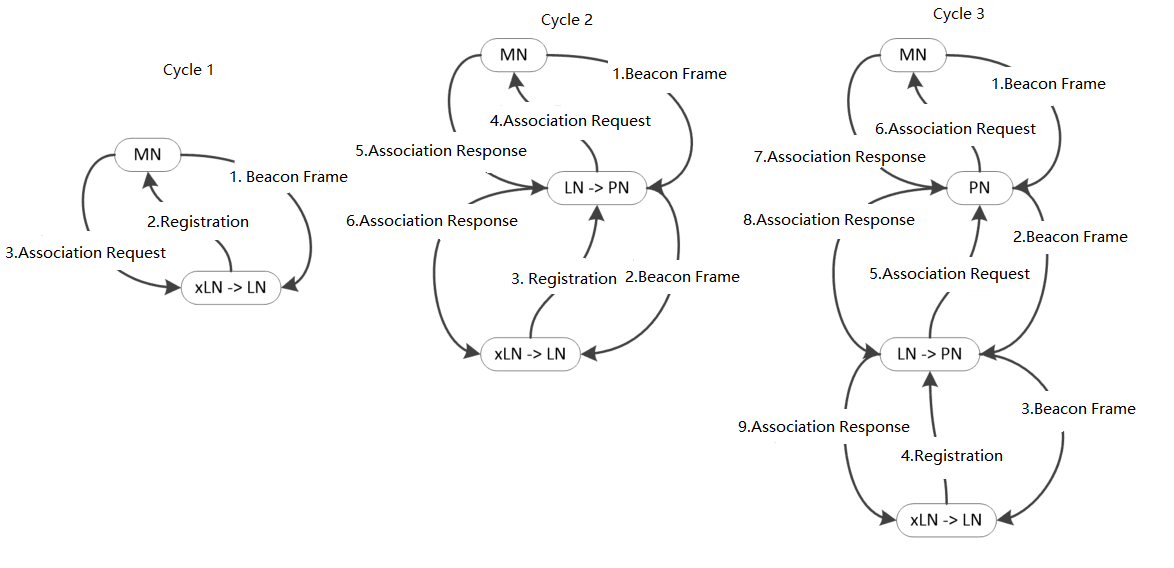}\\
	\caption{Frequency Division Power Line Network System Networking}
	\label{Figure_7}
\end{figure*}

\section{Simulation Results}
In this section, we evaluate our P-MAC mechanism in FD-PLC network in simulations, and compare it with the traditional CSMA.

The data packet is 12 OFDM symbols, each symbol has 1024 data, 64 cyclic prefixes. In the simulation, the sampling rate is 1.25 MSPS. The simulation uses a reasonable single preamble detection time slot, for example, the transmission time slot is 220 us. The preamble contention time slot is 256 time slots, which is 256 $\times$ 220 us, that is, 56.32 ms. For CSMA, the data packet contention time slot is set to maximum, i.e., 256 slots.

As shown in Fig. 9, the abscissa represents the number of nodes in FD-PLC network and the ordinate represents the time of establishing the network. For different frequency division power line network scales, compared with traditional CSMA access, the P-MAC mechanism can greatly shorten the time required for networking. For example, P-MAC requires 5.12s for 64 nodes, while CSMA requires 327.68s. PMAC reduces networking time by approximately 60 times compared with traditional CSMA access. P-MAC replaces data transmission with preamble frames for competitive access so that it greatly reduces the time of establishing the Ad-Hoc network.

\begin{figure}[!t]
	\centering
	\includegraphics[width=0.5\textwidth]{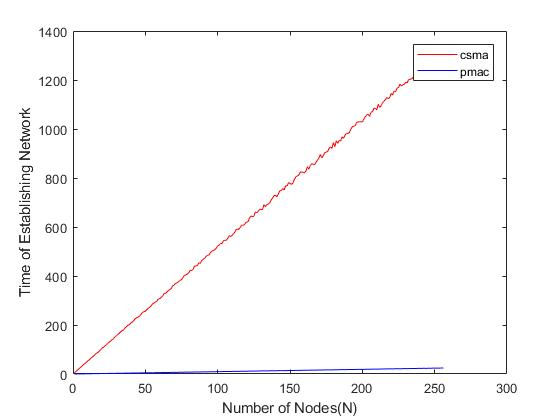}\\
	\caption{Time of Establishing Frequency Division Power Line Network System Networking}
	\label{Figure_8}
\end{figure}

The simulation also includes the network utilization rate when transmitting data after the completion of the frequency division power line network of different scales. Network utilization is the ratio of current network traffic to the maximum traffic that the port can handle. It indicates the bandwidth usage in the network. While high network utilization indicates the network is busy, and vice versa. As shown in Fig. 10, the abscissa represents the number of nodes in FD-PLC network and the ordinate represents the network utilization. The results of PMAC are similar to CSMA in terms of network utilization.

\begin{figure}[!t]
	\centering
	\includegraphics[width=0.5\textwidth]{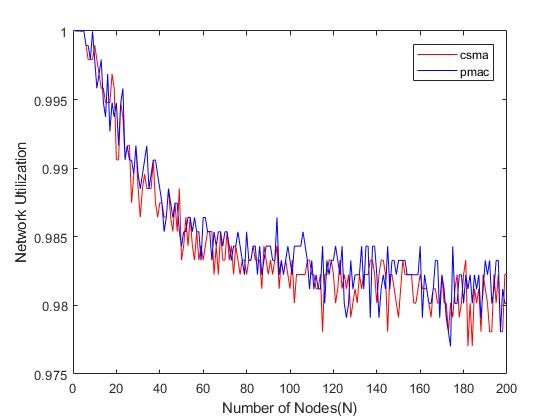}\\
	\caption{Network Utilization After the Completion of the Frequency Division Power Line Network}
	\label{Figure_9}
\end{figure}
The simulation includes the network utilization rate when establishing the FD-PLC network of different scales. As shown in Fig. 11, the abscissa represents the number of nodes in FD-PLC network and the ordinate represents the network utilization. The results of P-MAC are much better than those of CSMA in terms of network utilization.
\begin{figure}[!t]
	\centering
	\includegraphics[width=0.5\textwidth]{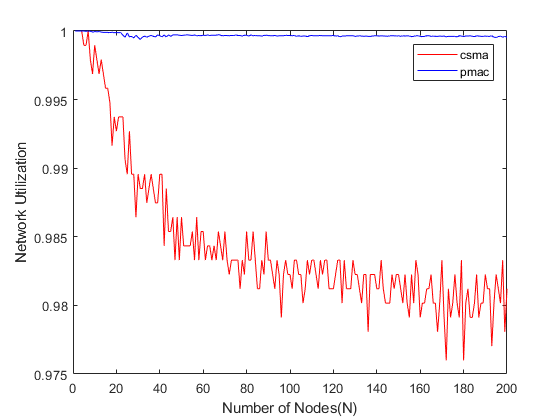}\\
	\caption{Network Utilization During the Frequency Division Power Line Networking}
	\label{Figure_10}
\end{figure}

\section{Conclusion}
In this paper, we propose the P-MAC mechanism in Ad-Hoc network to improve the networking efficiency. First, the P-MAC mechanism and maintenance network method is defined. Secondly, in order to optimize the Ad-Hoc PLC network we propose a FD-PLC network which can choose the best communication frequency. Next, due to the huge network scale in industrial scenarios, we combine FD-PLC and P-MAC to speed up the establishment of the Ad-Hoc PLC network. Finally, we evaluate our strategy and CSMA/CA for comparison. Numerical results indicates that P-MAC reduces networking time by approximately 60 times compared with traditional CSMA access. In the future, we expect to evaluate P-MAC under more practical FD-PLC network constraints. The main difficulty of implementing the P-MAC mechanism in practice will be the hardware implementation such as the design of cascade filters and the digital front end. The digital filter can be implemented in software or hardware. It is necessary to choose an implementation method that provides satisfactory performance under the condition of limited precision operation.


\section*{About the Authors}\footnotesize\vskip 2mm



\end{document}